# Title: A Pattern Discovery-Based Method for Detecting Multi-Locus Genetic Association

**Running Title: Pattern Discovery-based Association Test**

Zhong Li[1], Aris Floratos[3], David Wang[3], and Andrea Califano[2,3]

[1]Department of Computational Genetics, High Throughput Biology Inc., Livingston, NJ

[2]Department of Biomedical Informatics, Columbia University, New York, NY, USA

[3]Center for Computational Biology and Bioinformatics, Columbia University, New York, NY, USA

Address for correspondence and reprints: Dr. Zhong Li, 513 W. Mount Pleasant Ave, Suite 115, Livingston, NJ 07039. Telephone: 972-992-6222; Fax: 973-992-6225; E-mail: zli@htbiology.com

# Abstract

Methods to effectively detect multi-locus genetic association are becoming increasingly relevant in the genetic dissection of complex trait in humans. Current approaches typically consider a limited number of hypotheses, most of which are related to the effect of a single locus or of a relatively small number of neighboring loci on a chromosomal region. We have developed a novel method that is specifically designed to detect genetic association involving multiple disease-susceptibility loci, possibly on different chromosomes. Our approach relies on the efficient discovery of patterns comprising spatially unrestricted polymorphic markers and on the use of appropriate test statistics to evaluate pattern-trait association. Power calculations using multi-locus disease models demonstrate significant gain of power by using this method in detecting multi-locus genetic association when compared to a standard single marker analysis method. When analyzing a Schizophrenia dataset, we confirmed a previously identified gene-gene interaction. In addition, a less conspicuous association involving different markers on the same two genes was also identified, implicating genetic heterogeneity.

Keywords: Pattern, Power, Schizophrenia, Interaction

# Introduction

Genetic dissection of complex traits in humans has become a major focus in the biomedical research community [Lander, et al. 2001] as the majority of common diseases with significant public health impact fall in this category. It has been proposed that genome-wide association analysis may be the choice of deciphering the non-Mendelian inheritance mode of common diseases [Risch and Merikangas 1996]. However, up to now there have been limited successes with conventional association studies [Hugot, et al. 2001; Rioux, et al. 2001; Tavtigian, et al. 2001], possibly due to the multifactorial nature of common diseases. Multi-locus genetic association analysis seems to provide increased power to detect multiple susceptibility loci even with moderate sample size [Longmate 2001], and thus might be an advantageous alternative for the genetic dissection of complex traits in humans.

Methods for multi-locus association analysis can be classified into two broad categories: local multi-locus genetic association analysis and global multi-locus genetic association analysis [Hoh and Ott 2003]. Methods for local multi-locus genetic association analysis, such as the haplotype-based association test [Puffenberger, et al. 1994; Terwilliger and Ott 1992] and the haplotype-based Transmission Disequilibrium Test (TDT) [Clayton and Jones 1999; Clayton 1999; Dudbridge, et al. 2000; Lazzeroni and Lange 1998; Merriman, et al. 1998; Zhao, et al. 2000], usually analyze multiple nearby loci jointly. $\chi^2$ test of independence is often used to evaluate statistical significance of potential association [Spielman, et al. 1993; Terwilliger and Ott 1992].

More recently, a generalized $T^2$ test utilizing multiple markers was proposed as a more powerful alternative to $\chi^2$ test for association [Xiong, et al. 2002]. Methods for global multi-locus genetic association analysis [Hoh and Ott 2003], such as logistic regression [Cruickshanks, et al. 1992], sums of single-marker statistics [Hoh, et al. 2000], combinatorial partitioning [Nelson, et al. 2001], and data mining [Czika, et al. 2001; Flodman, et al. 2001; Toivonen, et al. 2000], focus instead on the detection of multiple disease susceptibility loci regardless of their chromosomal locations in the genome. Except for the combinatorial partitioning method, existing global multi-locus association analysis methods are not designed to systematically search all possible combination of loci in order to identify potential marker-trait association. As a result, a particular combination of marker loci with significant association to a trait might evade significance evaluation and go undetected.

Several challenges exist for developing such a method for global multi-locus association analysis. The foremost challenge arises from the combinatorial nature of the problem. Increasing the number of loci used in a genome-wide scan results in the exponential growth of possible multi-locus combinations, thus making the computational cost of their enumeration prohibitive. The second challenge is how to adjust statistical significance for multiple-testing. With tens of millions of multi-locus combinations evaluated and with strong correlations among different combinations due to locus sharing, proper multiple-testing corrections other than the Bonferroni correction are needed to control type I error while retaining power.

Pattern discovery-based approaches have been shown to be rather efficient at dealing with the combinatorial nature of identifying multi-element associations with analyses of protein motifs and gene expression data [Califano 2000; Rigoutsos and Floratos 1998]. Remarkably, a "Market Basket Analysis" (MBA) method, similar to the pattern discovery-based methods on sequence analysis and gene expression analysis, has been widely used in marketing research to identify association rules among disjointed sets of items [Agrawal and Srikant 1994].

We have developed a pattern discovery-based method to detect both local and global multi-locus genetic associations. The method employs an efficient pattern discovery algorithm and a pattern-trait association test to assess the statistical significance of potential associations. A model to correct pattern significance for multiple testing was also developed. In this report, we have applied the pattern discovery-based method to both simulated datasets and a real Schizophrenia dataset. Power calculations using multi-locus disease models on simulated genotypes on 71 SNP markers demonstrated significant gain of power by using this method in detecting multi-locus genetic associations when compared to a standard single marker analysis method. When analyzing a Schizophrenia dataset, we not only confirmed the original finding of a gene-gene interaction associated with Schizophrenia [Chumakov, et al. 2002], but also identified a potential association involving different alleles on the same two genes that had eluded previous detection efforts. Further analysis of this finding implies potential genetic heterogeneity. To summarize, our results indicated that the pattern discovery-based method for detecting multi-locus genetic association could be a useful tool for dissecting complex genetic traits in humans, particularly when oligogenic/polygenic inheritance and gene-gene interactions are present.

# Materials

## *Two-Locus Disease Models*

In this report, two-locus disease models [Clerget-Darpoux and Babron 1989; Dizier, et al. 1994; Fan and Knapp 2003; Xiong, et al. 2002] were chosen for power calculations (Table 1). Each disease-causing locus was represented by a bi-allelic marker (marker M1 for the first locus, marker M2 for the second locus). In the recessive-recessive (R-R) model, the presence of homozygote genotypes in both markers (*aa* for M1, *bb* for M2) was required to confer the disease state (genotype *aabb* has elevated disease penetrance). In the dominant-recessive (D-R) model, homozygote status was only required for marker M1 (both genotypes *aaBb* and *aabb* have elevated disease penetrance). To emulate the genetic complexity of common disease, we selected a high level of penetrance for phenocopies and modest penetrance for disease-causing genotype(s) [Gardner, et al. 1997; Lander and Schork 1994; Wallace, et al. 1996; Xiong, et al. 2002]. The penetrance of phenocopies (individuals that are affected but do not have the disease genotype(s) on M1 and M2) was set to 10%, while the penetrance of affected genotype was set to 26.4% for the D-R model and 44% for the R-R model.

## *Real Data*

Genotypes of 28 SNP markers on 454 individuals, including 213 affected individuals and 241 controls, were obtained from GENSET Corp. (now a subsidiary of Serono, S.A.). Of the 28 markers, 8 markers are on chromosome 12 at position 12q24, spanning 115 kb and covering the gene D-amino acid oxidase (DAO[MIM124050]) [Chumakov, et al. 2002].

The rest of the 20 markers are on 13q34, spanning 266 kb and covering the G72[MIM607408] gene [Chumakov, et al. 2002].

## *Simulated Data*

Simulated genotype data were used for power calculation. Genotype data (without phase information) on 71 SNP markers were modeled after an actual human population [Johnson, et al. 2001]. In particular, 69 SNP markers on five genes were picked to serve as disease-unrelated markers (13 markers for gene H19, 11 markers for CASP10, 13 markers for CASP8, 20 markers for SDF1, and 12 markers for TCF8). Two additional markers (M1 and M2) were embedded into each simulated genome, serving as the disease-related markers.

Assuming a population-based case-control study design, genotype data were generated *de novo* from a simulation program (Dahlia Nielsen, personal communication) by providing values for the following program parameters: haplotype/allele frequencies, penetrance of disease-causing genotypes, and penetrance of phenocopies. Phase information and haplotype structures were ignored in data analysis. For each of the two disease models tested, we evaluated five genotype frequency settings for disease-related markers M1 and M2 (Table 1) and two sample sizes (250 cases/250 controls, and 500 cases/500 controls). For each simulation condition, which was a combination of disease model, frequency setting, and sample size, 500 simulated datasets, each containing genotypes of 71 SNP markers, were generated. To derive test rejection level for individual pattern/marker, 500 simulated datasets were also generated for each simulation condition under the null hypothesis, in which the penetrance of phenocopies was used for all possible genotypes.

# Methods and Results

*Methods*

**The pattern discovery-based method for multi-locus genetic association analysis**

Detailed notations on pattern and maximal patterns can be found in Appendix. Briefly, a pattern is a sub-matrix $\pi$ of a data matrix $M$ (individuals as rows and markers as columns), identified by a subset of the markers $C_\pi$ (pattern composition) and a subset of individuals $S_\pi$ (pattern support), such that the value of each marker $X$ in $\pi$ across the individuals within $S_\pi$ satisfies a predefined equivalence criteria. Figure 1 showed two representative patterns (Figure 1B and 1C) embedded in a data matrix (Figure 1A). A sub-pattern is generated when one or more individuals (rows) and/or one or more markers (columns) are removed from a pattern. A maximal pattern is a pattern that is not a sub-pattern of any other pattern in the dataset. Given that a pattern is always more statistically significant than any of its sub-patterns [Califano 2000], only *maximal* patterns are identified in this method [Rigoutsos and Floratos 1998].

The pattern discovery-based method for association analysis was based on an efficient deterministic algorithm [Califano 2000], which ensures that all maximal patterns satisfying a single parameter, support constraint, will be discovered. The support constraint, $j_{min}$, specifies the minimum support size for any reported pattern, regardless of the pattern composition. The algorithm starts by first identifying all individual markers (pattern seeds) whose corresponding genotype(s) satisfy a specified $j_{min}$. Subsequently, pattern seeds are merged systematically in an iterative merging process, in which sub-patterns and patterns that are no longer satisfying the support constraint $j_{min}$ are removed, resulting in an exhaustive list of maximal patterns.

In the context of a case/control study, we seek maximal patterns that exhibit a significantly higher support in cases versus controls. Therefore, our method starts with a discovery step where maximal patterns are identified in cases only (Detailed explanation on this approach can be found in Appendix). Subsequently, each discovered pattern is evaluated through an association test statistics that accepts or rejects the null hypothesis of no distribution difference in the pattern support between cases and controls. As described in detail in the Appendix, the association test statistics for patterns is a goodness-of-fit test similar to the $\chi^2$ test. In the test, the presence or absence of a pattern in cases and controls are counted. As a result, only one degree of freedom is available for the test statistics. A reference distribution with only one degree of freedom that summarizes test statistics for all patterns regardless of the marker composition or support in a pattern, is generated with a Monte Carlo process, in which 1,000~10,000 simulated datasets are synthesized by randomly picking cases from the pool of actual cases and controls. The same Monte Carlo process is also used to implement a proper multiple testing correction for this method to control family-wise type I error.

**<u>Estimation of test rejection level for individual pattern/marker</u>**

The test rejection level for individual pattern or marker (the latter for the $\chi^2$-based single marker test) was estimated to achieve a family-wise type I error rate of 0.05. To study the correlation between power and the pattern support constraint $j_{min}$, test rejection level for individual pattern was established for every value of $j_{min}$. More specifically, at each value $j$ of $j_{min}$ within an arbitrary window, typically between 20% and 40% of the sample size for cases, the test rejection levels for individual pattern for a particular combination of sample size and genotype frequency were estimated in the following manner: First, pattern discovery was performed on each of the 500 datasets under the null hypothesis. Second, a reference distribution for pattern significance $D_j$ was constructed with test statistics obtained from patterns. Third, after pattern with the smallest P value was registered for each dataset, the smallest P values obtained from all 500 datasets were sorted in ascending order. Last, the largest P value within the top 5 percentile of the sort was chosen as the test rejection level ($\alpha_j$) for individual pattern for the specified value $j$ of $j_{min}$. The mean and standard deviation of the test rejection level was then estimated from all simulation conditions for a given sample size. An identical process (without the support stratification) was used to obtain the significance threshold $\alpha_0$ for the single-marker $\chi^2$ test.

**Power calculation**

Different procedures were used to estimate power for the pattern discovery-based method and for the single marker $\chi^2$ test. For the pattern discovery-based method, power was first calculated as a function of the pattern support constraint $j_{min}$. Specifically, under each simulation condition the following procedure was used: At each value $j$ of $j_{min}$ within an arbitrary window, typically between 20% and 40% of the case sample size, pattern discovery was performed on each of the 500 datasets. Resulting patterns identified at each $j$ were assigned P values using the corresponding reference distribution $D_j$ and were deemed "significant" if their P values were less than $\alpha_j$. A true positive call was registered for a dataset if the pattern with only the disease genotypes for markers M1 and M2 was found significant. Power was defined as the percentage of true positive calls for a given value $j$ of $j_{min}$ among all 500 datasets. The maximal power achieved under all values of $j_{min}$ was used as the representative power for the pattern discovery-based method under that particular simulation condition.

For the single marker $\chi^2$ test, the set of markers with P value less than $\alpha_0$ was selected. If both markers M1 and M2 were in that set, the test on that dataset was considered a true positive. The overall power of the single marker $\chi^2$ test was again defined as the percentage of the true positive calls.

## $\chi^2$-Based Single Marker Association Test

For the purpose of comparison, a $\chi^2$-based single marker association test was performed both to assess the statistical power of the approach based on simulated data and in the analysis of a Schizophrenia data set. In the test, a 2x2 table was created for each genotype of a marker, with the present or absent of the genotype in corresponding two rows and with case and control as two columns. $\chi^2$ statistics and corresponding P value were then obtained. (Please note: This $\chi^2$ test is slightly different from the one preformed in [Chumakov, et al. 2002]. In [Chumakov, et al. 2002], $\chi^2$ test was performed separately in French Canadian and Russian populations. In this analysis, $\chi^2$ test was performed on the combined population.)

## *Results*

**Estimation of Test Rejection Level for Individual Marker/Pattern**

Before estimating power, test rejection level for individual marker or pattern was established from genotype data simulated on 71 markers under the null hypothesis. The null hypothesis was simulated under two different sample sizes (250 cases/250 controls and 500 cases/500 controls) and ten genotype frequencies for marker M1 and M2 (five genotype frequencies as in the D-R model and five as in the R-R model). Test rejection levels for individual marker and pattern were obtained at each value $j$ of $j_{min}$ between 20% and 40% of the case sample size. Table 2 shows the mean test rejection levels for single marker and pattern with $j_{min}$ set to 20% of the case sample size. For the single marker $\chi^2$ test, test rejection levels are fairly consistent across different simulation conditions. The single marker test rejection levels under both sample sizes (P=1.01E-3 and P=1.08E-3 for 250 cases and 500 cases, respectively) were less conservative than the Bonferroni correction, which would yields a significance level of 7.2E-4, presumably due to haplotype structures embedded in the simulation model. Similar observations were made on the test rejection level for single pattern with $j_{min}$ set to 20% of the case sample size (P=1.19E-4 and P=1.05E-4 for 250 cases and 500 cases, respectively). With on average 4,000 patterns identified from each simulated dataset with $j_{min}$ set to 20% of the case sample size, Bonferroni correction would have yielded a significant threshold of 1.25E-5.

**Power Calculation and Comparison**

A simulation-based power calculation was performed to evaluate the power of the pattern discovery-based method on multi-locus association analysis. For comparison purposes, the power of a $\chi^2$-based single marker test was also analyzed. Power was analyzed with genotypes of 71 SNP markers for two selected disease models (Dominant-Recessive and Recessive-Recessive) as a function of the sample size and the genotype frequencies of affected markers M1 and M2. Because the power of the pattern discovery-based method may vary when pattern discovery is performed with different pattern support constraint $j_{min}$, the correlation between power and $j_{min}$ was analyzed prior to the power comparison.

Figure 2 shows the correlation between power and pattern support constraint $j_{min}$ under 20 different simulation conditions, with each condition representing a particular combination of sample size, disease model, and frequency of affected genotypes. Correlations with 250 cases and 250 controls are shown in Figure 2A, while correlations with sample size of 500 cases and 500 controls are shown in Figure 2B. As expected, a general trend of power reduction was observed with the increase of $j_{min}$ under most simulation conditions, although the onset and the magnitude of power reduction varied under different conditions (Figure 2). The onset of power reduction is strongly correlated with the frequency of affected genotypes as higher frequency leads to a larger proportion of affected individuals that share the affected genotypes.

No significant power difference was observed between the two disease models, suggesting that the pattern discovery-based method might perform similarly across genetic models, which is advantageous since the underline genetic models for common diseases are normally not known.
13

While under many simulation conditions the power has demonstrated a monotonic decrease with the increase of $j_{min}$ (For example: Figure 2B, conditions with affected genotype frequency of 0.143, 0.206, and 0.28 for the D-R disease model, and conditions with affected genotype frequency of 0.08, 0.115, and 0.156 for the R-R disease model), a power peak was nevertheless evident under the dominant-recessive model (Figure 2A, conditions with affected genotype frequency of 0.143, 0.206, and 0.28). For example, with affected genotype frequency of 0.206 and 250 cases under the dominant-recessive model, the maximal power (87%) was achieved with $j_{optimal}$=72 (Table 3, row 4). The fact that a loss of power was observed with $j_{min}$ smaller than the $j_{optimal}$ confirmed our hypothesis that setting $j_{min}$ smaller than an empirically estimated $j_{optimal}$ for the pattern discovery step of the method is detrimental, because a smaller $j_{min}$ would yield a larger number of total patterns, which in turn would lead to lower test rejection level for individual pattern due to multiple testing correction.

The optimal support constraint $j_{optimal}$ was identified for all simulation conditions with $j_{min}$ varying between 20% and 40% of the case sample size (Table 3, Figure 2). Because no clear power peak was visible for two of the smallest frequencies (5.1% and 9.1% for D-R, 2.9% and 5.1% for R-R) of affected genotypes under both disease models and both sample sizes, we further reduced $j_{min}$ to 10% of the case sample size to examine the possibility that larger power can be recovered with smaller $j_{min}$. Indeed larger power was obtained in all eight simulation conditions, resulting in new $j_{optimal}$ for each condition (Table 3, rows 1, 2, 6, and 7). The best power identified under each simulation condition (Table 3) was used for power comparison between the pattern discovery-based method for association analysis and the single marker $\chi^2$ test (Figure 3).

Under all 20 simulation conditions, the pattern discovery-based methods consistently outperformed the single marker $\chi^2$ test in terms of absolute power (Figure 3A and 3C). While power for both methods improved with the larger sample size (500 vs. 250) and higher affected genotype frequency, the gain of power by the pattern discovery-based method over the single marker association test is not uniform (Figure 3B and 3D). Instead, the pattern discovery-based method exhibited the largest gain of power when the affected genotype frequency was the smallest. For example, a more than 10 fold power increase over the single marker association test at affected genotype frequency of 5.1% was observed for the pattern discovery-based method for the D-R disease model with 500 cases and 500 controls (Figure 3B). In the R-R disease model, a 27 fold power increase at affected genotype frequency of 2.9% was observed with 250 cases and 250 controls. On the other hand, there were less than 2 fold increases of power by the pattern discovery-based method over the single marker association test with affected genotype frequencies of 28% for the D-R model and of 15.6% for the R-R model. Differential power gain by the pattern discovery-based method over the single marker test was also observed at two sample sizes tested. As shown in Figure 3B and 3D, except for the simulation condition with affected genotype frequency of 5.1% in the D-R model, larger power ratios between these two methods were consistently observed with the sample size of 250 cases and 250 controls in comparison with the sample size of 500 cases and 500 controls. Taking together, these observations indicate that the pattern discovery-based method is able to detect disease association under all designed simulation conditions and might have significant advantage over the traditional single marker association test when the affected genotype frequency is small.

**Detecting Gene-Gene Interactions in a Schizophrenia Dataset**

In order to further assess the utility of the pattern discovery-based method for association analysis, we applied it to a well-characterized dataset collected from a case/control association study on Schizophrenia. The dataset contained genotypes for 28 SNP markers spanning 115 kb at 12q24 (8 markers) and 266 kb at 13q34 (20 markers). Using traditional association tests and functional assays, two interacting susceptibility genes were previously identified on these regions of the human genome [Chumakov, et al. 2002]: G72 on chromosome 13, and D-amino acid oxidase (DAO) on chromosome 12. Specifically, markers M-22 on chromosome 13 and MDAAO-6 on chromosome 12 were found to show significant association with the disease when considered together. Because markers in this dataset are distributed over two different chromosomal regions and the associated markers are known, this dataset is ideal to test the pattern discovery-based method for multi-locus genetic association.

To establish a baseline, we used a frequency-based single-marker $\chi^2$ test to detect genotype distribution bias in the dataset, similar to what was described [Chumakov, et al. 2002]. Consistent with previous findings [Chumakov, et al. 2002], three markers were found to be statistically significant (P <= 0.01) with the single marker association test before multiple-testing correction (MDAAO-6: P=9.62e-4; M-22: P=5.84e-3; MDAAO-5: P=7.71e-3). With Bonferroni correction for multiple-testing to achieve a family-wise type I error of 0.05 (test rejection level for individual marker is 0.00183), however, only MDAAO-6 remained significant.

The pattern discovery-based method was applied to both the real dataset and 5,000 simulated datasets. Because the true statistical power for this method on this real dataset is unknown, we used the number of significant patterns as a representation of power to estimate the value of the optimal support constraint $j_{optimal}$ (the Appendix). $j_{optimal}$ was estimated to be 24, roughly 10% of the case population under study (Figure 4). With the association test rejection level for individual pattern calculated at 9.49e-5 with $j_{min} = j_{optimal}$, only two out of 5,952 identified patterns were found to be significantly associated with Schizophrenia. None of the sub-patterns of those two significant patterns were significantly associated with Schizophrenia.

The most significant pattern (Pattern A in Table 4A) included two markers, marker MDAAO-6 on chromosome 12, and M-22 on chromosome 13, the same two markers that were previously identified in a multi-locus association with Schizophrenia [Chumakov, et al. 2002]. Relative risk analysis using a Crude Odds Ratio test [Hennekens and Buring 1987] on pattern A provided additional support to the significance of the association (Table 4A, odds ratio=4.54, confidence interval: 2.3-9.1). The second significant pattern (Pattern B in Table 4A) included seven markers, one marker (marker MDAAO-7) on chromosome 12, and markers M-11, M-12, M-13, M-14, M-15, and M-16 (M-11~M-16) on chromosome 13. None of the markers in this pattern was significant by itself when assessed with the single-marker $\chi^2$ test (Table 4A). The significance of the association was only found when they were considered together in a single pattern (P=7.33e-5). Again, crude odds ratio analysis indicated that pattern B carried significant relative risk for Schizophrenia (odds ratio=15, 95% confidence internal: 3.5-65).

Despite the fact that marker M-22 and markers M-11~M-16 are co-localized in a 266 kb region on chromosome 13, they belong to different LD blocks: Markers M-12~M-16 form a single LD block (D' between markers >0.9), while M-22 is in a LD block with M-23 and M-24 [Chumakov, et al. 2002]. Similarly, though barely 9.7 kb apart, markers MDAAO-6 and MDAAO-7 are not in strong linkage disequilibrium (D'=0.55). Combined with the fact that markers in both patterns were mapped to the same two genes (DAO and G72), these observations prompted us to ask the following question: Do those two patterns indicate independent patterns of co-segregation, perhaps due to genetic heterogeneity? If that was the case, one would expect that the affected individuals carrying the MDAAO-6/M-22 mutations would be relatively disjoint from the affected individuals carrying the M-11~M-16/MDAAO-7 mutations. Indeed, when individuals participated in pattern A and B are tallied, only three cases and none of the controls participated in both patterns (Table 4B). To calculate the combined relative risk on DAO and G72 loci as characterized by Pattern A or Pattern B for Schizophrenia, we constructed a $\chi^2$ test (Table 4C) in which individuals carrying either of the two patterns or none of the two patterns were counted in cases and in controls. A significant P value (P=8.67E-11) suggested that at least two independent patterns of co-segregation might be responsible for the susceptibility on the DAO and G72 loci.

# Discussion

The pattern discovery-based method for multi-locus genetic association has several novel properties. First, unlike current local multi-locus analysis methods [Lazzeroni and Lange 1998; Puffenberger, et al. 1994; Terwilliger and Ott 1992], which are usually restricted to nearby markers, a pattern can contain markers located far apart on a chromosome or even on different chromosomes. It is, therefore, ideally suited for the identification of all possible co-segregations in the genome, which might be a powerful way to effectively locate multiple disease susceptibility genes responsible for common diseases. Second, comparing to other global multi-locus analysis methods such as the logistic regression analysis [Cruickshanks, et al. 1992] and the sum of single-marker statistics [Hoh, et al. 2000], our method does not rely on single-locus effects. Instead, it can detect both single-locus and multi-locus effects simultaneously. Third, the deterministic nature of the algorithm guarantees the discovery of all maximal patterns, thus reducing the risk of missing important multi-locus association. The identification of the second significant association involving seven markers from the Schizophrenia dataset by the pattern discovery-based method supports this notion. Fourth, this approach is model-independent and can be applied to various study designs such as case/control and family-based studies, making it a general method for genetic analysis.

      One of the most significant challenges for the pattern discovery-based method is how to correct for multiple testing. For example, even a moderate sized dataset with a few hundred markers can produce tens of thousands of highly correlated patterns,

Bonferroni correction is clearly not appropriate for it. Instead, we have employed a Monte Carlo process [Lazzeroni and Lange 1998; McIntyre, et al. 2000] to empirically estimate the appropriate significance threshold for the single pattern association test so that the family-wise type I error can be controlled to an acceptable level. Furthermore, by estimating the significance threshold for single pattern at each $j_{min}$, our method further limits the detrimental effect of random patterns with small support, as demonstrated in power calculation, where the power of our method actually decreased when $j_{min}$ was less than $j_{optimal}$ under several simulation conditions.

Under two disease models we demonstrated that our pattern discovery-based multi-locus association test has superior power to detect multi-locus association when compared to a single marker association test. To properly generate simulation datasets, several assumptions reflecting the underlying genetic complexity of common diseases were implemented, including factors such as incomplete penetrance, polygenic inheritance, and high frequency of disease-causing alleles, etc.. In both disease models, polygenic inheritance was required for modestly elevated but partial disease penetrance. A rather substantial (10%) penetrance for phenocopies was used. On those two markers representing the susceptibility loci, both affected alleles were common alleles with high allele frequencies (from 23% to 75%, depending on the disease model and simulation condition). By demonstrating strong power under these simulation conditions, we argue that the pattern discovery-based method is a valid approach for detecting associations of genetic factors to complex traits. Furthermore, simulations suggested that the pattern discovery-based method gained most power with relatively small sample sizes and low allele frequencies for associated allele.

The discovery of two significant patterns from the Schizophrenia dataset provided a good example to illustrate some of the advantages of using the pattern discovery-based method. Not only did our method confirmed original finding of a gene-gene interaction (Table 4, pattern A) [Chumakov, et al. 2002], it also detected a potential association (Table 4, pattern B) that was not identified previously. Interestingly, none of the markers in pattern B was significant individually, only by analyzing all possible combinations of these markers were we able to identify the potential association. It is critical to confirm the genetic association suggested by pattern B with an independent dataset in future studies and with functional assays.

In summary, our analysis suggested that the pattern discovery-based method for multi-locus genetic association analysis is a potentially powerful test that can be applied to the dissection of complex genetic traits in humans. When compared to a standard single marker association test, the pattern discovery based method appears to have more power at detecting susceptibility loci, especially when multiple disease-causing genes are present. In order to use the method in genome-wide association studies, the efficiency of the pattern discovery algorithm needs to be improved so that tens of thousands of markers can be readily accommodated in a single search. At the present, a divide-and-conquer approach is designed to handle large dataset collected from genome-wide association studies. With proper heuristics, we hope to be able to reduce the search space for maximal patterns while retaining power.

# Appendix

## *Definitions*

Genotypic data for I individuals on N genomic markers can be arranged in a matrix $M = \{\mu_{i,m}\}$, where each row $1 \leq i \leq I$ corresponds to an individual and each column $1 \leq m \leq N$ corresponds to a marker (SNP or microsatellite, for example). Cell $\mu_{i,m}$ contains the genotype of the *m*-th marker on the *i*-th individual. Henceforth, a pattern on this data set shall be defined as a sub-matrix $\pi$ of *M*, identified by a subset of the markers $C_\pi$ (pattern composition) and a subset of individuals $S_\pi$ (pattern support), such that the value of each marker *X* in $\pi$ across the individuals within $S_\pi$ satisfies a predefined equivalence criteria. In this work we use "identity" (i.e., the marker values for *X* must be identical across all the individuals contained in the pattern support set) as the equivalence criteria, although more complex equivalence criteria are also possible (e.g., for dealing with microsatellite markers).

A pattern $\pi$ is called a *jk* pattern if $|S_\pi| = j$ and $|C_\pi| = k$. Such a pattern is said to "have support *j*". Depending on the nature of the genomic regions used for each individual (single or paired chromosomes), a pattern can be further categorized as *allelic* or *genotypic*. Allelic patterns are made up of chromosomal alleles, while genotypic patterns comprise di-allelic genotypes. An allelic pattern whose constituent markers are co-located on the same chromosome is equivalent to a haplotype. The definition of a pattern is illustrated in Figure 1 with a data matrix of eight individuals genotyped at nine single nucleotide polymorphisms (SNPs). Two genotypic patterns are shown. The first one (A, shown with italics on a green background in Figures 1A and 1B) has composition $C_A = \{M_1, M_4, M_8\}$ and support $S_A = \{I_3, I_4, I_6\}$. The second pattern (B, shown as boldface on a yellow background in Figures 1A and 1C) has composition $C_B = \{M_3, M_5, M_6, M_9\}$ and support $S_B = \{I_1, I_2, I_5, I_7, I_8\}$.

Removal from a pattern of one or more individuals (rows) and/or one or more markers (columns) will result in a new sub-matrix that is also a pattern. The latter will be called a sub-pattern of the former. For a pattern, the number of its possible sub-patterns is exponential both in the size of its support and its composition. Given that a pattern is always more statistically significant than any of its sub-patterns [Califano 2000], it is desirable to detect only *maximal* patterns [Rigoutsos and Floratos 1998]. A pattern is deemed maximal if it is not a sub-pattern of any other pattern in the dataset. For instance, both patterns A and B in Fig. 1 are maximal. However pattern C, with support $S_C = \{I_3, I_6\}$ and composition $C_C = \{M_1 M_4\}$, is not maximal because it is a sub-pattern of B obtained by removing $I_4$ from its support and $M_8$ from its composition.

## *A Pattern Discovery-Based Multi-Locus Association Test*

Many pattern discovery algorithms are available both of a deterministic [Califano 2000; Jonassen 1997; Rigoutsos and Floratos 1998] and statistical [Bailey and Elkan 1994; Neuwald and Green 1994] nature. In this paper, we adapted one of the most efficient deterministic algorithms [Califano 2000] for use on genotypic data. The deterministic nature of the algorithm guarantees that all maximal patterns satisfying two parameters, support constraint and composition constraint, will be discovered. The support constraint, $j_{min}$, specifies the minimum support size for any reported pattern. The composition constraint, $k_{min}$, specifies the minimum size of composition for any reported pattern.

Because their values impact overall running time, $k_{min}$ and $j_{min}$ should be set prudently, balancing the need to discover as many significant patterns as possible with tolerance for computational performance. In the analysis of genotypic data, we set $k_{min}$ to 1 in order to capture single-marker patterns, effectively leaving the support constraint as the only parameter for the pattern discovery algorithm. The value of the support constraint, $j_{min}$, is empirically estimated, taking into account study-related and performance-related issues.

Maximal patterns are identified in the following algorithmic steps: (1) The formation of seed pattern. A seed pattern is a maximal pattern satisfying $j_{min}$ with only one marker. (2) The formation of maximal pattern. Seed patterns are merged to form maximal patterns satisfying $j_{min}$ with more than one marker. Sub-patterns and patterns that no longer satisfy $j_{min}$ are removed from the merging process. By managing the merging process based on similarities among seed patterns, our algorithm rapidly identifies all qualifying maximal patterns without exploring all possible combinations among seed patterns. The optimality of such procedure has been described [Califano 2000].

In the context of a case/control study, we are seeking maximal patterns that exhibit a significantly higher support in the case versus the control group. Therefore, our method commences by a discovery step where maximal patterns are only identified in cases (the case-only approach), which is followed by significance evaluation through an association test statistics that accept or reject the null hypothesis of no distribution difference in the pattern support between cases and controls. Comparing to an alternative approach in which patterns are discovered in cases plus controls and then evaluated in cases against controls (the case+control approach), the case-only approach has the following advantages: (1) Better performance. The smaller sample size of the case-only approach results in less maximal patterns and less computation. (2) Lower false negative rate. As shown in the Results section, the power of this pattern discovery-based association test is directly correlated with the pattern support constraint. Because the pattern support constraint for case-only patterns can be set much lower than the support constraint for case-control patterns due to the smaller sample size in case-only approach, better power can be achieved with the case-only approach. (3) Broader maximal pattern coverage. It can be shown that while all maximal patterns identified in the cases are maximal in cases+controls, not all maximal patterns identified in cases+controls are maximal in cases only.

In the test statistic, a 2x2 contingency table is constructed with one column for cases and one for controls. Cells in the first row of the table register supports of a pattern in cases and in controls. Cells in the second row register the numbers of cases and controls that are genotyped on markers in a pattern but do not contain the pattern. Alternatively, a 2xn contingency table can also be used with n-1 degree of freedom and with n observed genotype/allele combinations among markers in a pattern. A goodness-of-fit test similar to the $\chi^2$ test is then performed. P values are derived from a reference distribution that is generated with a Monte Carlo permutation process. In the Monte Carlo process, typically 1,000~10,000 simulated datasets are synthesized by randomly picking cases from the pool of actual cases and controls. Pattern discovery is then performed on each simulated dataset with the same constraint parameters $k_{min}$ and $j_{min}$ used in actual dataset. A reference distribution is subsequently generated from tallies of test statistics obtained from each simulation.

One serious challenge for our approach is the multiple-testing problem. Even a moderate sized dataset with just a few hundred markers can contain tens of thousands of maximal patterns, many of which can be spurious associations. Although Bonferroni correction is routinely used to estimate the test rejection level for individual marker ($\alpha_i$) so that the family-wise type I error ($\alpha$) is controlled (normally to 5%), it is too conservative for patterns (as shown below in Results), because of extensive correlations among patterns due to marker sharing.

To properly estimate $\alpha_i$, we again used outputs obtained from the Monte Carlo simulation process from which the reference distribution for test statistics is produced for a target dataset. Assuming the null hypothesis for all of the simulated datasets, a family-wise type I error rate of 0.05 means that significant patterns appear in no more than 5% of the simulated datasets. Patterns were discovered within each simulated dataset and were assigned P values according to the reference distribution discussed above. The smallest P value from each dataset was ordered again in ascending order for all simulated datasets. The largest P value at the top 5 percentile was declared the test rejection level for individual pattern.

In summary, on a dataset collected from a case-control study, the pattern discovery-based association test containing the following steps: (1) Maximal patterns are identified from the case dataset with properly chosen $j_{min}$; (2) Maximal patterns are evaluated for significant association using a goodness-of-fit statistical test under the null hypothesis that both cases and controls are from the same sample population. Pattern significance is adjusted for multiple-testing with simulation.

## *Estimating* $j_{optimal}$

Selecting an appropriate $j_{min}$ for pattern discovery is an interesting problem of its own right. While choosing a large value for $j_{min}$ can significantly reduce the computation required to identify maximal patterns, it can also adversely affect the power of the pattern discovery-based association test: a large $j_{min}$ value can cause true associations with smaller representations in the sampled population to be missed. On the other hand, choosing a very small $j_{min}$ can be detrimental as well: computationally, the number of maximal patterns increases exponentially as $j_{min}$ decreases; statistically, multiple testing correction drives the threshold for type I error substantially lower with a small $j_{min}$, thus rendering many patterns insignificant, including (potentially) some patterns that may correspond to real disease associations. The choosing of $j_{min}$ is also related to the study design, from which the dataset is collected. Consider, for example, a case/control study involving 100 cases and 100 controls with a hypothetical 0.0056 threshold for test rejection level for individual pattern. Assume further that pattern significance is to be evaluated using a $\chi^2$ test comparing the support of the pattern in the case population versus the control population. If a pattern is expected to appear by chance alone in at least 1 control, then that pattern would need to have a support level of at least 10 within the case group in order to clear the 0.0056 significance threshold. Under these assumptions, if we were to use the pattern discovery-based method to discover patterns within the case group, it would make little sense to set $j_{min}$ to a value less than 10: Doing so would be detrimental as it would only increase the amount of computation required without adding any new significant patterns.

When the underlying disease model is known, simulations can be used to identify a value $j_{optimal}$ (the optimal $j_{min}$) that maximizes power because correlations between $j_{min}$ and power can be systematically established, as shown in Results. However, when the disease model is unknown, power cannot be formally established by simulation. In such cases, other criteria have to be used to estimate the $j_{optimal}$. For the Schizophrenia dataset, we empirically estimate $j_{optimal}$ as the support level for which the highest number of significant patterns is found. This heuristic is designed to maximize the number of candidate associations between genetic factors and the complex trait under investigation.

At each value $j$ of $j_{min}$ within an arbitrary window, typically between 5% and 40% of the case sample size, the test rejection level for individual pattern and reference distribution ($\alpha_j$ and $D_j$, respectively) are established with a Monte Carlo process. The selection of the arbitrary window is based on the following two factors: (1) the minimal expected support of a true association ($j_{min}$ = 5% of the case sample size means that the smallest representation of a true association our method can detect is 5% of the case population), and (2) computational efficiency. At each value $j$ of $j_{min}$ within the window, patterns are identified from the real dataset and evaluated against the corresponding reference distribution. $j_{optimal}$ is the largest $j_{min}$ to yield the largest number of significant patterns.

# Electronic-Database Information

Online Mendelian Inheritance in Man (OMIM), http://www.ncbi.nlm.nih.gov/Omim/

# Acknowledgment

The authors would like to thank Drs. Daniel Cohen and Fabio Macciardi for providing the GENSET dataset and valuable discussions. We would like to thank Dr. Dahlia Nielsen for her simulation program and stimulating discussions. We would also thank Dr. Aravinda Chakravarti for many helpful discussions. This work is partially supported by a SBIR phase I grant awarded by NCI (1R43CA101432-01).

# Tables

**Table 1** Disease models in power calculation.

| Model | Bi-allele Frequency | | | | LD (D') | Frequency of Affected Genotype(s) | Population Prevalence |
| --- | --- | --- | --- | --- | --- | --- | --- |
| | *A-B* | *A-b* | *a-B* | *a-b* | | | |
| D-R | 0.485 | 0.168 | 0.262 | 0.085 | 0.08 | 0.051 | 0.108 |
| D-R | 0.4 | 0.139 | 0.349 | 0.113 | 0.12 | 0.091 | 0.115 |
| D-R | 0.314 | 0.109 | 0.436 | 0.141 | 0.14 | 0.143 | 0.123 |
| D-R | 0.228 | 0.079 | 0.523 | 0.169 | 0.17 | 0.206 | 0.134 |
| D-R | 0.143 | 0.05 | 0.61 | 0.197 | 0.15 | 0.28 | 0.146 |
| R-R | 0.24 | 0.37 | 0.221 | 0.169 | 0.21 | 0.029 | 0.110 |
| R-R | 0.224 | 0.345 | 0.206 | 0.226 | 0.24 | 0.051 | 0.117 |
| R-R | 0.207 | 0.32 | 0.191 | 0.282 | 0.21 | 0.08 | 0.127 |
| R-R | 0.191 | 0.295 | 0.176 | 0.339 | 0.19 | 0.115 | 0.139 |
| R-R | 0.175 | 0.269 | 0.161 | 0.395 | 0.19 | 0.156 | 0.153 |

D-R represents the dominant-recessive disease model; R-R represents the recessive-recessive disease model. *A-B*, *A-b*, *a-B*, and *a-b* represent four possible allele combinations for markers M1 (alleles *A* and *a*) and M2 (alleles *B* and *b*). D' was calculated as described [Devlin and Risch 1995]. "Frequency of Affected Genotype(s)" represents the cumulative frequency of disease-causing genotype(s). For the D-R model, the disease-causing genotypes are *aaBb* and *aabb*. For the R-R model, the disease-causing genotype is *aabb*.

**Table 2** Estimates of significance thresholds (individual type-I error) used in power calculations.

| Sample Size | Single Marker Goodness-of-fit | | Pattern Discovery-Based | |
|---|---|---|---|---|
| | Mean | SD | Mean | SD |
| 250 cases/250 controls | 1.01E-3 | 1.63E-4 | 1.19e-4 | 2.60e-5 |
| 500 cases/500 controls | 1.08E-3 | 2.29E-4 | 1.05e-4 | 4.82e-5 |

Thresholds are chosen so that the family-wise type I error is 0.05. Values in the "Mean" column were obtained by averaging the rejection levels from each of 10 datasets generated using the simulation conditions described in the text. For the pattern discovery-based method, the values displayed are for support constraint $j_{min}$ equal to 20% of the size of the case group.

**Table 3** $j_{optimal}$ in power calculation.

| Disease Model | Genotype Frequency | 250 cases/250 controls | | 500 cases/500 controls | |
|---|---|---|---|---|---|
| | | $j_{optimal}$ | Power | $j_{optimal}$ | Power |
| D-R | 0.051 | 26 | 0.04 | 50 | 0.27 |
| D-R | 0.091 | 46 | 0.24 | 71 | 0.71 |
| D-R | 0.143 | 56 | 0.60 | 109 | 0.99 |
| D-R | 0.206 | 72 | 0.87 | 148 | 1 |
| D-R | 0.28 | 90 | 0.98 | 182 | 1 |
| R-R | 0.029 | 25 | 0.34 | 50 | 0.72 |
| R-R | 0.051 | 38 | 0.81 | 71 | 0.89 |
| R-R | 0.08 | 50 | 1 | 108 | 1 |
| R-R | 0.115 | 72 | 1 | 150 | 0.99 |
| R-R | 0.156 | 85 | 1 | 187 | 1 |

**Table 4** Gene-gene interactions and genetic heterogeneity detected by two patterns in GENSET dataset (only markers in significant pattern were shown).

**A**

| Marker | Chromosome | Genotype | Location (base) | P value | Pattern A | Pattern B |
|---|---|---|---|---|---|---|
| M-11 | 13 | C/C | 88,511,524 | 3.59E-2 | | x |
| M-12 | 13 | A/A | 88,535,455 | 1.28E-2 | | x |
| M-13 | 13 | C/C | 88,542,989 | 7.36E-2 | | x |
| M-14 | 13 | G/G | 88,549,446 | 4.55E-2 | | x |
| M-15 | 13 | A/A | 88,551,544 | 3.59E-2 | | x |
| M-16 | 13 | G/G | 88,552,517 | 2.06E-1 | | x |
| M-22 | 13 | G/G | 88,601,302 | 5.84E-3 | x | |
| MDAAO-6 | 12 | T/T | 109,018,154 | 9.6E-4 | x | |
| MDAAO-7 | 12 | A/A | 109,027,872 | 2.26E-2 | | x |
| | | | | | | |
| Case Support | | | | | 38 | 24 |
| Control Support | | | | | 11 | 2 |
| P value | | | | | 4.64E-5 | 7.33E-5 |
| Odds Ratio | | | | | 4.54 | 15.17 |
| Confidence Int. | | | | | 2.2-9.1 | 3.5-65.0 |

The "x" marks indicate the marker composition for a pattern.

**B**

| | Pattern A Only | Pattern A and B | Pattern B only |
|---|---|---|---|
| Support in Cases | 35 | 3 | 21 |
| Support in Controls | 11 | 0 | 2 |

**C**

| | Case | Control |
|---|---|---|
| Individuals with either Pattern A or B | 59 | 13 |
| Individuals with neither Pattern A nor B | 154 | 228 |
| | P value=8.67E-11 | |

# Figures

**Figure 1** Examples of maximal patterns.

A sample genotype data matrix with 9 SNP markers (M1~M9) on 8 individuals (I1~I8) was shown. Genotype is represented as two alleles. Cells corresponding to pattern A or B were color-coded.

A. Data Matrix

|    | M1  | M2  | M3  | M4  | M5  | M6  | M7  | M8  | M9  |
|----|-----|-----|-----|-----|-----|-----|-----|-----|-----|
| I1 | A/A | T/T | **A/A** | T/T | **C/G** | **T/T** | A/T | G/G | **T/T** |
| I2 | G/G | A/T | **A/A** | C/C | **C/G** | **T/T** | A/A | G/G | **T/T** |
| I3 | *A/G* | A/A | T/T | *T/C* | C/C | C/T | T/T | *T/G* | T/G |
| I4 | *A/G* | A/T | A/T | *T/C* | G/G | C/C | A/T | *T/G* | T/G |
| I5 | G/G | A/A | **A/A** | T/C | **C/G** | **T/T** | A/A | T/G | **T/T** |
| I6 | *A/G* | A/T | A/A | *T/C* | C/C | C/T | T/T | *T/G* | G/G |
| I7 | A/G | A/A | **A/A** | T/T | **C/G** | **T/T** | A/A | T/T | **T/T** |
| I8 | G/G | A/A | **A/A** | T/C | **C/G** | **T/T** | A/T | T/T | **T/T** |

B. Pattern A

|    | M1  | M4  | M8  |
|----|-----|-----|-----|
| I3 | *A/G* | *T/C* | *T/G* |
| I4 | *A/G* | *T/C* | *T/G* |
| I6 | *A/G* | *T/C* | *T/G* |

C. Pattern B

|    | M3  | M5  | M6  | M9  |
|----|-----|-----|-----|-----|
| I1 | **A/A** | **C/G** | **T/T** | **T/T** |
| I2 | **A/A** | **C/G** | **T/T** | **T/T** |
| I5 | **A/A** | **C/G** | **T/T** | **T/T** |
| I7 | **A/A** | **C/G** | **T/T** | **T/T** |
| I8 | **A/A** | **C/G** | **T/T** | **T/T** |

**Figure 2** Correlation between power and pattern support constraint.

**A**. Power was calculated at each pattern support constraint $j_{min}$ between 50 and 100 under ten different simulation conditions with sample size of 250 cases and 250 controls.

**B**. Power was calculated at each pattern support constraint $j_{min}$ between 100 and 200 under ten different simulation conditions with sample size of 500 cases and 500 controls. Filled "x" represents affected genotype frequency of 0.051 under the dominant-recessive model. Filled circles represent affected genotype frequency of 0.091 under the dominant-recessive model. Filled diamonds represent affected genotype frequency of 0.143 under the dominant-recessive model. Filled triangles represent affected genotype frequency of 0.206 under the dominant-recessive model. Filled squares represent affected genotype frequency of 0.28 under the dominant-recessive model. "x" represents affected genotype frequency of 0.029 under the recessive-recessive model. Unfilled circles represent affected genotype frequency of 0.051 under the recessive-recessive model. Unfilled diamonds represent affected genotype frequency of 0.08 under the recessive-recessive model. Unfilled triangles represent affected genotype frequency of 0.115 under the recessive-recessive model. Unfilled squares represent affected genotype frequency of 0.156 under the recessive-recessive model.

A

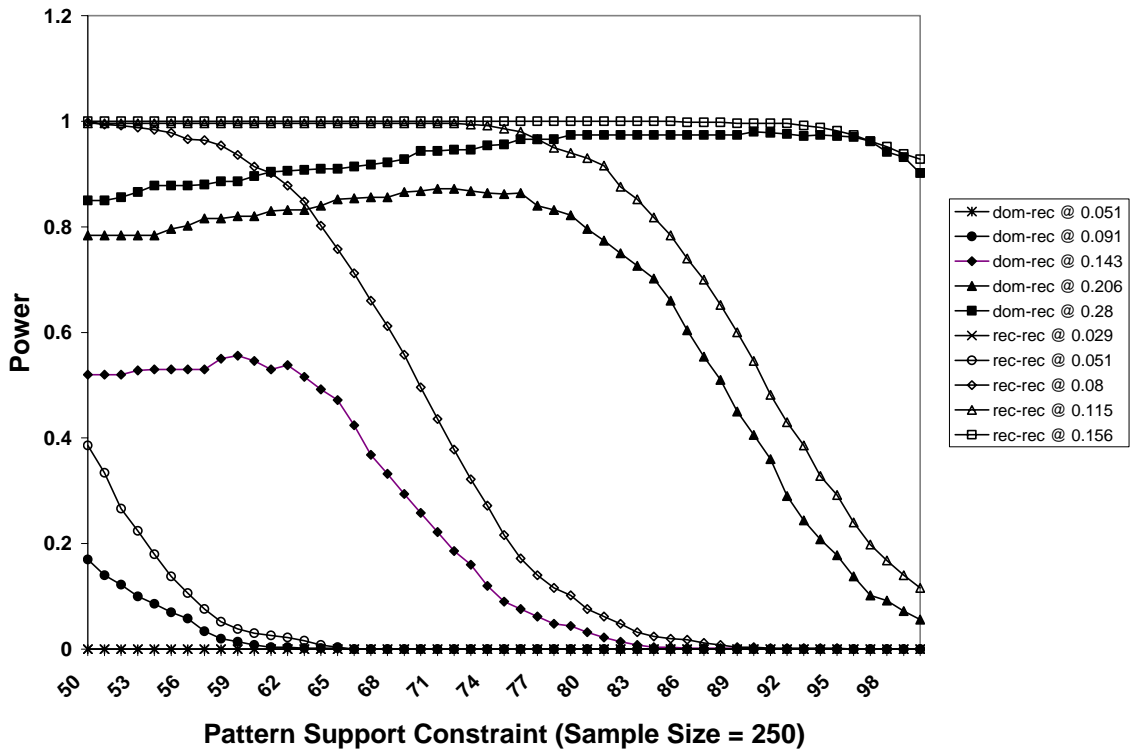

B

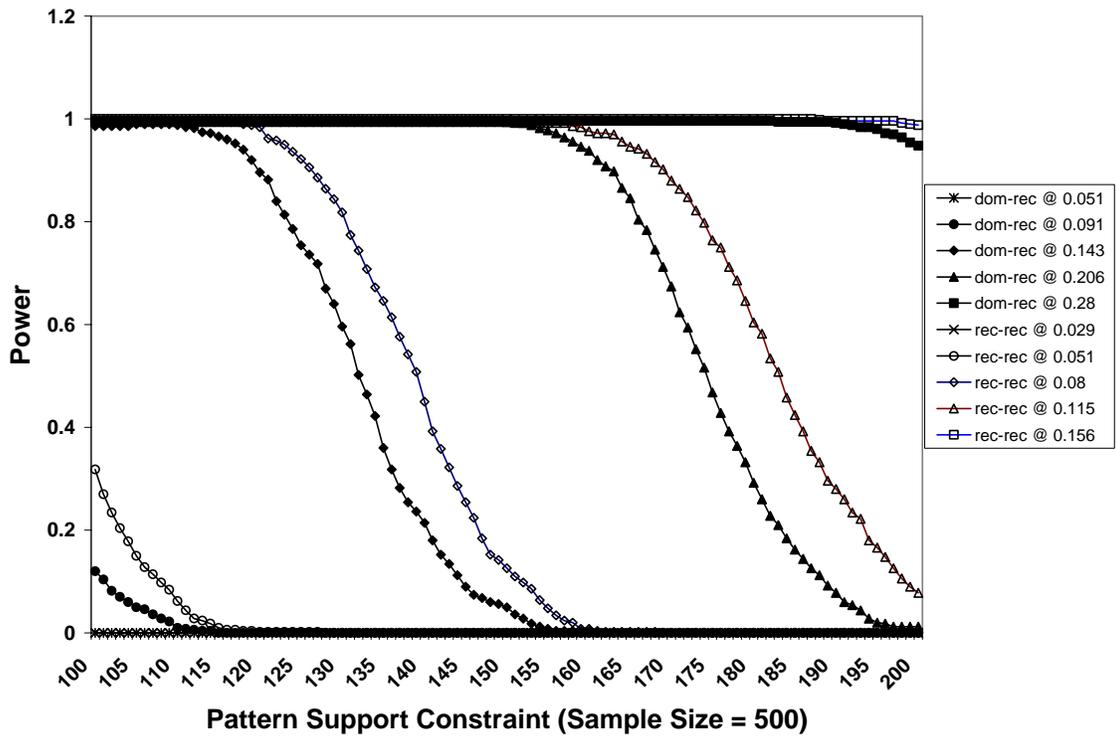

**Figure 3** Power comparisons between a single marker $\chi^2$ test vs. the pattern discovery-based method.

Powers were compared between a single marker $\chi^2$ test and the pattern discovery-based multi-locus association test under two disease models (dominant-recessive in A and B, recessive-recessive in C and D). A and C showed the power curves for both methods with two sample sizes (250 cases/250 controls and 500 cases/500 controls). B and D showed the power ratio between the pattern discovery-based method and the single marker $\chi^2$ test under corresponding sample sizes.

**A** 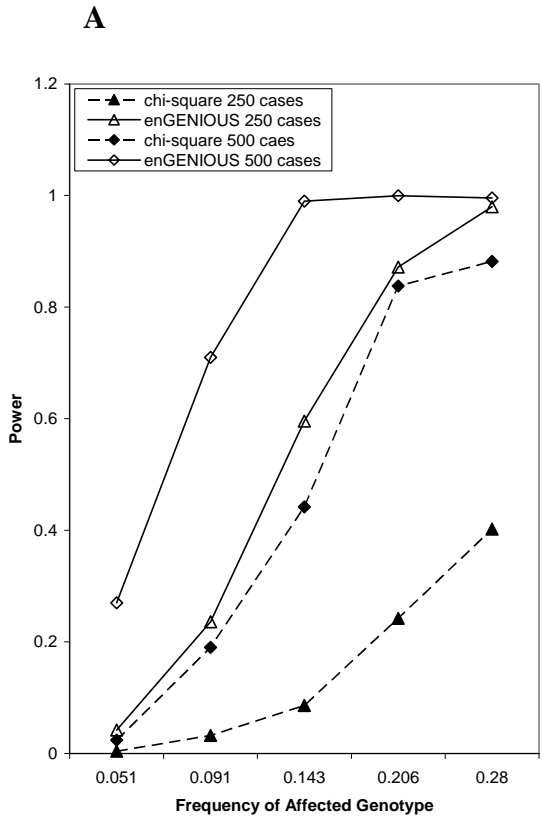　**B** 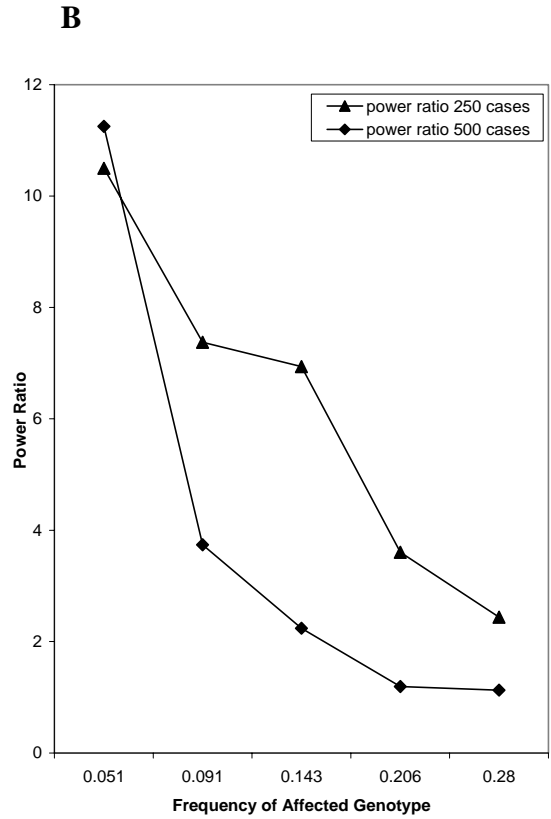

**C**                        **D**

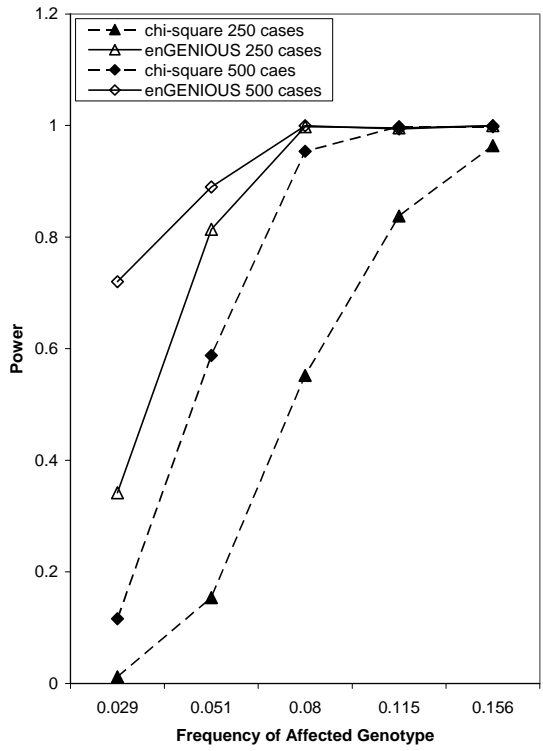
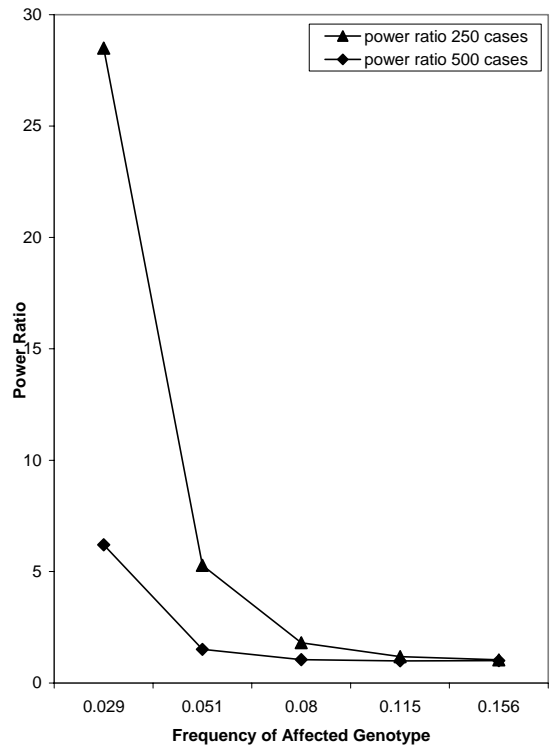

**Figure 4** Estimating the optimal support constraint for the GENSET dataset. Significant patterns were identified with the pattern discovery-based method at pattern support constraint $j_{min}$ between 15 and 50 from the GENSET dataset. The number of significant patterns at each $j_{min}$ was plotted against the $j_{min}$. The optimal support constraint (24) is the largest $j_{min}$ with the largest number of significant patterns.

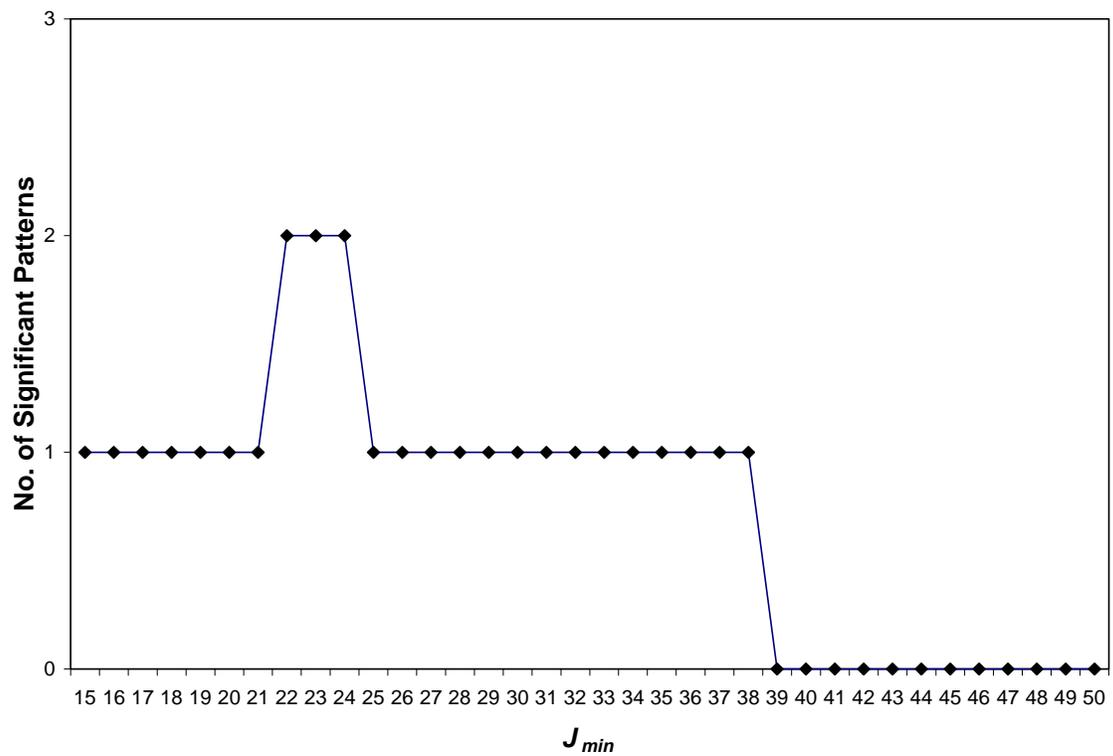